\documentclass[preprint,showpacs,aps,prl]{revtex4}
\usepackage{epsfig}
\usepackage{bm}
\usepackage{latexsym}

\begin{document}
\title{Phase diagram for diblock copolymer melts 
\\ under cylindrical confinement}

\author{Weihua Li and Robert A. Wickham}
\affiliation{Department of Physics, St.\ Francis Xavier University, 
Antigonish, Nova Scotia, Canada B2G 2W5}

\date{\today}

\begin{abstract}
We extensively study the phase diagram of a diblock copolymer melt
confined in a cylindrical nanopore
 using real-space self-consistent mean-field theory. 
We discover a rich variety of new two-dimensional 
equilibrium structures that have no 
analog in the unconfined system.
These include 
non-hexagonally coordinated cylinder phases and structures intermediate 
between lamellae and cylinders.  We map the stability 
regions and phase boundaries for all the structures we find.
As the 
pore radius is decreased, the pore accommodates fewer cylindrical 
domains and structural
transitions occur as cylinders are eliminated. Our results are consistent 
with experiments, but we also predict phases yet to be observed.

\end{abstract}

\pacs{82.35.Jk, 61.41.+e, 61.46.+w}

\maketitle

Self-assembly in macromolecular systems provides a convenient route to create
structure at the nanoscale. These structures have potential applications 
as, for example, lithographic templates for nanowires, photonic crystals, 
and high-density magnetic storage media \cite{PARK03}. Confinement effects, 
produced by boundaries, influence the self-assembly process and can
generate novel nanostructures. In a rational search for new 
morphologies to fit a particular application, it is crucial to 
know which structures are possible in a confined system, 
and under what conditions, {\em i.e.}\ one 
needs to know the phase diagram.

Confined diblock copolymer melts have become the focus of increasing attention.
In the bulk, these materials self-assemble into a variety of periodic
nanostructures --- lamellae, hexagonally-coordinated cylinders, body-centred 
cubic lattices of spheres, and the gyroid morphology. The ability to tune the
period and to control self-assembly using 
temperature, 
chemical composition, and molecular architecture, lend a rich physics to 
these materials, and make them attractive for industrial applications 
\cite{PARK03,PARK97,SEGALMAN01,CHENG04}. 
Competition between the chain stretching energy and the interfacial energy 
between block domains determines which structures form in the bulk.
Introducing confinement modifies this competition. 
Confinement of diblock copolymers between parallel solid walls, or in
 a thin film, has been extensively 
studied \cite{MATSEN97,MORKVED97,GEISINGER99,HUININK00,RASMUSSEN04}. 
However, the effect of confinement in a cylindrical pore 
is relatively unexplored. 

Recently, Xiang {\em et al.}\ examined asymmetric 
and symmetric 
polystyrene-{\em b}-polybutadiene (PS-{\em b}-PBD) diblock copolymers 
confined to cylindrical alumina nanopores and observed cylindrical 
and concentric lamellar structures in the pores \cite{XIANG04}. 
Concentric lamellar 
structures were also seen in the experiments of Sun {\em et al.}\ 
involving symmetric polystyrene-{\em b}-poly(methyl methacrylate) 
(PS-{\em b}-PMMA) diblock copolymers confined in alumina nanopores
\cite{SUN05}.
Concentric lamellar structures have been seen in the Monte Carlo 
simulations of  He {\em et al.}\ \cite{HE01} and in the dynamical 
density functional simulations of Sevink {\em et al.}\ 
\cite{SEVINK01}. Experiments using narrower pores find that stacked-disc or 
toroidal-type structures, and also cylindrical helices, are possible 
\cite{SHIN04,XIANG05}. Such three-dimensional structures have been seen in 
the simulations of Ref.\ \cite{SEVINK01} and in a self-consistent 
mean-field theory (SCMFT) study of self-assembled silica-surfactant 
mesostructures in cylindrical nanopores \cite{WU04}. However, in
all the simulation and theoretical work to date, only a few phases, in a very 
limited region of parameter space, have been studied due to the time and
computational requirements. Knowledge of the phase
behaviour of cylindrically confined block copolymers is 
thus currently fragmentary.

To fill this knowledge gap, we systematically evaluate 
the phase diagram for a diblock copolymer melt confined in a cylindrical 
nanopore using SCMFT. SCMFT
has proved to be one of the most successful theoretical methods for 
investigating equilibrium phases in block copolymers, and has played a
major role in establishing the phase diagram of the bulk diblock copolymer 
melt \cite{MATSEN94,MATSEN02}. We explore the 
phase diagram for the cylindrically confined melt, and identify 
the equilibrium phase boundaries and stability regions for various 
phases that exist in this system. We consider structures that are
translationally-invariant along the pore axis, and are thus two-dimensional.
Previous works have 
examined on the order of 10 points in the phase diagram. Our study examines 
approximately 3000 points in the phase diagram and consumes
 approximately 15,000 cpu-hours of a 3 GHz Intel Xeon processor. 

%
%
We consider an incompressible melt of AB diblock copolymers, confined to a
cylindrical pore of radius $R$. Each copolymer has a degree of 
polymerization $N$ while the A-block on each has a degree of polymerization 
$fN$ with $0\le f \le 1$. Lengths in our theory are expressed in units of 
the radius of gyration, $R_g$, of the polymer.
Within the mean-field approximation to the 
many-chain Edwards theory
\cite{HELFAND75,HONG81}, at a temperature $T$ the free-energy $F$ per chain
for $n$ Gaussian diblock copolymer chains confined in a cylindrical pore 
has the form 
%
%
\begin{eqnarray}
\label{EQ:FE}
\frac{F}{nk_{B}T} & = & -\ln Q + \frac{1}{V}\int_{|{\bf r}| \le R} 
d{\bf r} \{  \chi N
\phi_{A}({\bf r})\phi_{B}({\bf r}) - \omega_{A}({\bf r}) \phi_{A}({\bf r})
- \omega_{B}({\bf r}) \phi_{B}({\bf r}) 
\nonumber \\ & &
\hspace{3 in} 
+ H({\bf r})[ \phi_{A}({\bf r})
-\phi_{B}({\bf r})] \}.
\end{eqnarray}
The monomer densities are $\phi_A$ and $\phi_B$, the partition 
function $Q$ is 
for a single polymer interacting with the mean-fields $\omega_{A}$ 
and $\omega_{B}$ produced by the surrounding chains. These quantities 
have the standard definitions and meanings \cite{MATSEN94,MATSEN02}. The 
Flory-Huggins interaction parameter, $\chi$, characterizes 
the repulsion between dissimilar monomers. In the 
confined melt, the spatial integration is restricted to the cylinder 
volume, taken to be $V$. We assume the pore wall has a preference for B 
monomers and include a surface field $H({\bf r})$ in Eq.\ (\ref{EQ:FE}) 
that attracts species B.
For convenience, this surface field is chosen to have the form
%
%
\begin{equation}
\frac{H({\bf r})}{\chi N} = V_{0}\{\exp[(\sigma+|{\bf r}|-R)/\lambda]-1\}
\label{EQ:SURF}
\end{equation} 
for $R-\sigma \le |{\bf r}| \le R$, while  $H({\bf r}) = 0$ for 
$|{\bf r}| < R-\sigma$. We choose the cutoff distance for the 
surface interaction to be  $\sigma=0.4$ $R_g$, and the decay length to 
be $\lambda = 0.2$ $R_g$. 
We fix the strength of the surface field 
to be $V_0=0.4$. We find that modest variations in $V_0$ have little effect 
on the phase diagram and morphologies.

Minimization of the free-energy with respect to the monomer densities and 
mean-fields leads to a set of mean-field equations which can be solved 
self-consistently in real space to find the equilibrium densities 
\cite{MATSEN94,MATSEN02,DROLET99}.
We employ the split-step Fourier method of Tzeremes {\em et al.}
\cite{TZEREMES02} to solve the modified diffusion equations for the 
end-segment distribution functions \cite{MATSEN94} on a $128\times128$ 
square lattice, with a lattice constant of $0.136$ $R_g$. 
The chain contour length for each block is 
discretized into 128 segments. We find that the stability regions and phase
boundaries are not significantly altered if we use a finer mesh, indicating
that a $128\times128$ lattice provides sufficient accuracy \cite{ACCURACY}.
Incompressibility, $\phi_A ({\bf r}) + \phi_B({\bf r})=1$, is enforced {\em 
via} a Lagrange multiplier for $|{\bf r}|<R$. Outside the pore, 
$|{\bf r}|\ge R$, there is no polymer, so we set the end-segment distribution
functions to zero outside the pore, 
which implies that $\phi_A=\phi_B=0$ in this region. 
Other than assuming that the 
structures are translation-invariant along the axis of the 
pore, we do not make 
any {\em a priori} assumptions about the symmetry of potential equilibrium
structures.
This is a major advantage of the real-space approach \cite{DROLET99}.
We first use random initial conditions in our iterative algorithm to generate 
a large set of solutions to the mean-field equations 
over a wide region of phase space.  Once we determine this set, we then use 
these solutions as initial conditions in our algorithm, to explore the extent 
(if any) of their stability regions. We take the 
equilibrium phase to be the structure which has the 
lowest free-energy, for a given $f$ and $\chi N$, of all the observed 
structures.
%
%

We first examine the phases that form in a fixed pore radius, 
$R=8.5$ $R_g$, and later discuss results for different radii. 
The 21 equilibrium 
nanostructured phases found at this pore radius are shown in 
Fig.\ \ref{FIG:STRUCT}.
 The notation we use for the structures is explained in the caption
of Fig.\ \ref{FIG:STRUCT}.
The phase diagram is presented in Fig.\
\ref{FIG:PHASE}, and has some features in common with the phase
diagram for the unconfined diblock copolymer melt \cite{MATSEN94}. 
The disordered (D) phase is 
stable when $\chi N$ is low, or when the copolymer is close to being a 
homopolymer ($f\rightarrow0$ or 1). Above the order-disorder 
transition, for $f \stackrel{<}{_\sim} 0.3$, the asymmetry in the block 
composition favours curved interfaces and the formation of cylindrical 
domains of the minority A species. Inverted cylindrical domains, composed of 
the minority B species, form when $f \stackrel{>}{_\sim} 0.65$. Symmetric 
copolymers favour AB interfaces with low curvature and thus concentric
lamellar rings form in the centre of the phase diagram.
There are, however, significant differences between our phase diagram 
and the phase diagram for the bulk system, which we now discuss.

Our phase diagram is much
more intricate than the phase diagram for the unconfined melt. Many of the
trends in the phase diagram can be understood in terms of packing a finite
number of domains into the pore, which has a finite cross-sectional area.
For a given $R$, $f$ and $\chi N$, there is an optimum number of domains that
can be accommodated --- this number minimizes the free-energy penalties 
associated with chain stretching or compression. At fixed $R$, as $\chi N$ 
increases, we see from Figs.\ \ref{FIG:STRUCT} and \ref{FIG:PHASE} 
that the domain size increases, and the number of domains in the pore 
decreases. For example, as $\chi N$ increases at fixed $f = 0.22$ the
structure changes from $C_{1-6-13}$ to $C_{1-6-11}$ to $C_{1-5-11}$
to $C_{4-10}$. A similar trend is seen for the inverted cylinder phases,
and for the concentric lamellae. For the cylinder phases of minority A (B) 
species, a similar argument explains the trends with increasing (decreasing)
$f$. These packing considerations are purely the result of confinement, and 
do not arise in the bulk.

We observe arrangements of cylinders in the inner region of the 
pore with  non-hexagonal coordination,  such as the 
$C_{4-10}$, $C_{1-5-11}$ an $\bar{C}_{1-7}$ phases. Cylindrical structures 
with non-hexagonal coordination are not seen as equilibrium 
phases in the bulk, since such structures produce a highly
 non-uniform majority domain thickness, with an associated free-energy
penalty 
\cite{MATSEN02}. 
Our observation of non-hexagonally coordinated
cylinders suggests that, when confined, the system will tolerate 
a non-uniform majority domain thickness
if it can avoid the compression (or stretching)
penalty of packing a non-optimum number of cylinders into the pore.

For a narrow region $0.3 \stackrel{<}{_\sim} f \stackrel{<}{_\sim}  0.4$
 in Fig.\ \ref{FIG:PHASE} we see structures that are intermediate between 
cylinders and lamellae. As in the unconfined melt, a value of $f$ closer to 
0.5 favours AB interfaces of lower curvature. For example, for 
$\chi N = 23$, as $f$ increases above $f\approx0.3$ the $C_{4-10}$ 
structure will transform into $LC_4$ as the presence of the pore wall 
first leads the outer ring of cylinders to coalesce into a single 
narrow domain of 
the A phase. With increasing $f$, the AB interfacial curvature is lowered 
further as cylinders coalesce into lamellar segments 
in the inner region and the system transforms
to the $LC_{L,2}$ and then $LC_{LL}$ phases. When $f$ increases above 
$f \approx 0.4$ the high curvature of the semi-circular end-caps on the 
lamellar segments becomes unfavorable, leading to the formation of concentric 
lamellar rings. Like the gyroid phase in the unconfined diblock copolymer 
melt, these intermediate phases represent a compromise between 
pure cylindrical and pure lamellar order.

There is an asymmetry in the phase diagram, arising from the 
preference of the pore wall for the B block.
In  Figs.\ \ref{FIG:STRUCT} and \ref{FIG:PHASE} the arrangements of the 
inverted cylinders and the 
stability regions for these structures are less complicated than the 
cylindrical structures seen for $f \stackrel{<}{_\sim} 0.3$. 
One can think about the behaviour in this region in terms of the 
behaviour for $f \stackrel{<}{_\sim} 0.3$ by exchanging the 
$A$ and $B$ monomers, 
and recognizing that the inverted cylinders 
exist in a pore with a smaller effective radius. In the 
inverted cylinder phases, a thick double layer of A 
monomers exists between the thin layer of minority 
B block along the pore wall and 
the cylinders in the inner region. We suggest that the outer A block,
attached to the B block along the actual pore wall, acts as filler, creating
an effective pore wall (of smaller radius) with a preference for A.
As we see below, when the radius of the pore decreases, fewer 
cylinders form in the pore.

 All the order-order transitions we observe are first-order phase transitions. 
As the order-disorder transition (ODT) 
is approached the inner region can disorder first 
(the $C_{1-6-13}$ to $C_{0-0-13}$ transition, for example) before the 
final transition to disorder. 
The ODT is first order, except perhaps for the ODT from $L_3$ to $D$. We 
are unable to determine the location of the ODT from $L_3$ to $D$,
and the dashed curve shown in this region in Fig.\ \ref{FIG:PHASE} is an 
interpolation of the full ODT curve. As the $L_3$ to $D$ transition is 
approached along the ODT, from high or low $f$, 
the first order ODT becomes weaker.
We speculate that 
the $L_3$ to $D$ transition may be continuous or, possibly, a crossover.

We explore the effect of changing the pore radius $R$ for the point
$f=0.26$ and $\chi N = 20$. Our results are shown in Fig.\ \ref{FIG:RADIUS}.
With increasing pore radius, the complexity of the structures, both in terms 
of number of cylinders and the number of rings, increases. We observe the
effect of commensurability of the pore radius with the hexagonal arrangement
of cylinders expected in the bulk --- the $C_{1-6}$ phase appears in smaller 
pores and then, following intervening, non-hexagonal phases, reappears as the 
$C_{1-6-12}$ phase in larger pores. Our expectation is that, for larger 
pore sizes, a hexagonal array of cylinders will preferentially form in the 
inner region of the pore in order to minimize unfavourable 
non-uniformities in the majority domain thickness. 
The $C_{1-6}$, $C_{1-7}$ and $C_{1-8}$ phases we observe here correspond, when 
A is exchanged with B, to structures seen in Fig.\ \ref{FIG:PHASE} for 
$f \stackrel{>}{_\sim} 0.65$, supporting the idea that these latter 
structures form due to confinement in a pore with a smaller effective
size.

Experiments by Xiang {\em et al.} examined both symmetric and
asymmetric PS-{\it b}-PBD diblock copolymer melts  confined to cylindrical 
nanopores in an alumina membrane \cite{XIANG04}. The PBD block 
preferentially segregated to the 
pore wall and can therefore be identified as the 
B block. Using an asymmetric copolymer, with a PS volume fraction of 0.64, 
they observed
a structure similar to $\bar{C}_{1-6}$ in a pore of approximate 
diameter $140$ nm (Fig.\ 4 of Ref.\ \cite{XIANG04}) . Using a symmetric 
copolymer, with a PS volume fraction of 0.44, 
they observed two concentric lamellar rings, similar to 
our $L_2$ structure, in an approximately $170$ nm 
diameter pore (Fig.\ 6 of Ref.\ \cite{XIANG04}). We estimate that 
$R_g \approx 7-8$ nm in these experiments; therefore 
the $17$ $R_g$ pore diameter used here is comparable the experimental 
pore diameter. The experiments of Sun {\em et al.}, involving symmetric 
PS-{\em b}-PMMA ($R_g \approx 10$ nm), also observe the $L_2$ phase in 180 nm 
diameter pores \cite{SUN05}. Structures with non-hexagonal symmetry, such as
the ones we predict here, have not been observed experimentally, to our
knowledge. Our work suggests that such structures exist for smaller
values of $f$ or higher temperatures than used in experiments. Recent 
experiments suggest that three-dimensional structures can form in narrow 
pores (of radius 2--4 $R_g$) \cite{SHIN04,XIANG05}; furthermore, we 
expect spherical domains to form when the composition of the copolymer
is highly asymmetric. We are currently investigating how our phase diagram
is modified when three-dimensional structures are allowed in the theory.

%
%

To summarize, we have extensively 
explored the phase diagram of a diblock copolymer melt confined to a 
cylindrical nanopore through the use of real-space SCMFT. The phase 
diagram is more complicated than for the unconfined system, and we have found 
new two-dimensional structures not seen in the bulk, notably
non-hexagonally coordinated cylindrical phases and structures intermediate 
between lamellae and cylinders.  A major achievement of our work is 
that our map of the phase stability regions and
phase transition curves now enables
thermodynamic issues for this 
system, such as the nature of transitions between different structures, to 
be explored.
Many trends in the phase diagram can be understood in terms of 
the ability of the system to pack domains 
into the finite-radius pore. For example, we observe that with 
decreasing pore radius fewer cylinders can be accommodated.
Our results are consistent with recent experimental observations, but we also
predict a wealth of phases in parameter 
regions yet to be explored experimentally.

%
%
%
\begin{acknowledgements}

The authors gratefully acknowledge helpful discussions with Profs. A.-C.\ Shi, 
D.\ Hunter, and P.\ H.\ Poole, and with Dr.\ K.\ Rasmussen. The authors also 
thank the StFX hpcLAB
and G.\ Lukeman for computing resources and support. This work was 
supported by NSERC, CFI and AIF. 

\end{acknowledgements}
%

%
%
\pagebreak
\begin{center}
{\bf Figure Captions }
\end{center}
%
%
\begin{figure}[!ht]
\caption{
Monomer density plots of the 21 nanostructured phases 
formed in the $8.5$  $R_g$ radius pore. The colour ranges from 
deep red (A-rich regions) to deep blue (B-rich regions). The region 
outside the cylindrical pore is also 
coloured deep blue. We use the notation $C$ for cylindrical phases,
 $L$ for lamellar 
phases, and $LC$ for structures containing both lamellae and cylinders,
which we call
``intermediate phases''. When the cylinders are composed of minority B 
component
an over-bar ($\bar{C}$ or $L\bar{C}$)  is used. The number of cylinders in each
ring, out from the centre of the pore, is indicated by subscripts  
${\rm C}_{i-j-k}$. The number of $L$ subscripts in the notation 
${\rm L C}_{L \cdots L, i-j}$ indicates the number lamellar segments 
 in the inner region of a given intermediate
structure. The second subscript 
indicates the number of cylinders of the minority A species 
in the pore, and whether these cylinders are arranged in rings.
The stability regions for these
structures are labelled on the phase diagram in Figure \ref{FIG:PHASE}.
}
\label{FIG:STRUCT}
\end{figure}
%
%
\begin{figure}[!ht]
\caption{Phase diagram for a diblock copolymer melt confined in a cylindrical 
nanopore of radius $R=8.5$ $R_g$. The degree of polymerization of the copolymer
is $N$, the Flory-Huggins parameter is $\chi$, and $f$ is the $A$ monomer
fraction. The disordered phase is labelled $D$. The dashed curve is an  
interpolation of the order-disorder transition curve, as explained in the 
text.}
\label{FIG:PHASE}
\end{figure}
%
%
%
\begin{figure}[!ht]
\caption{Phase stability regions as a function of pore radius. 
Colours indicate the stability range for each of 
the structures indicated,
as the radius of the the cylindrical pore is increased from $5$ $R_g$ to 
$9$ $R_g$ at $f=0.26$ and $\chi N=20$.}
\label{FIG:RADIUS}
\end{figure}
\pagebreak
%
%
\begin{figure}[h]
\epsfig{file=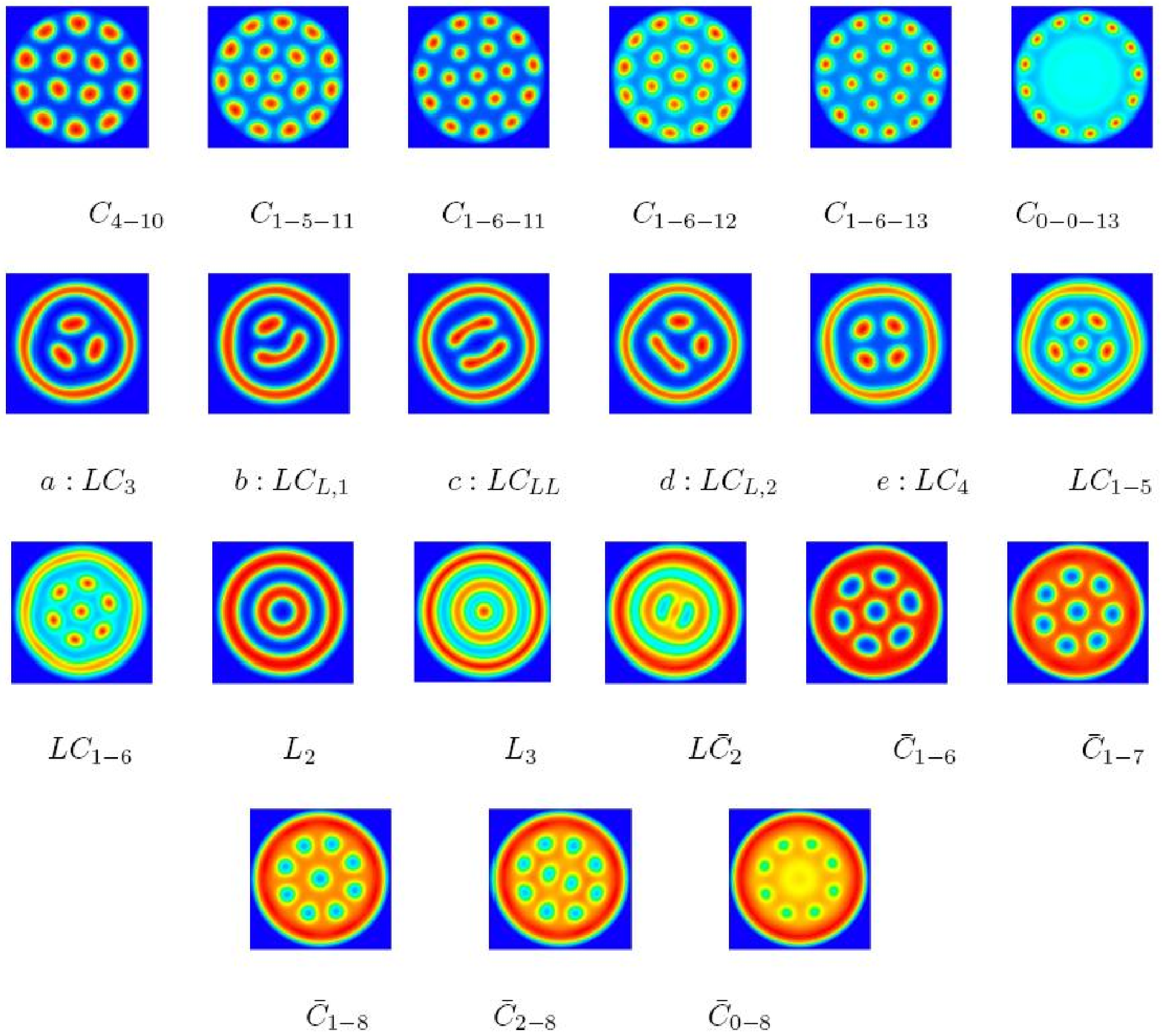, width=0.9\linewidth} \hspace{.3\linewidth}
\\ {\bf Figure 1 --- Li}
\end{figure}
\pagebreak
\begin{figure}[h]
\epsfig{file=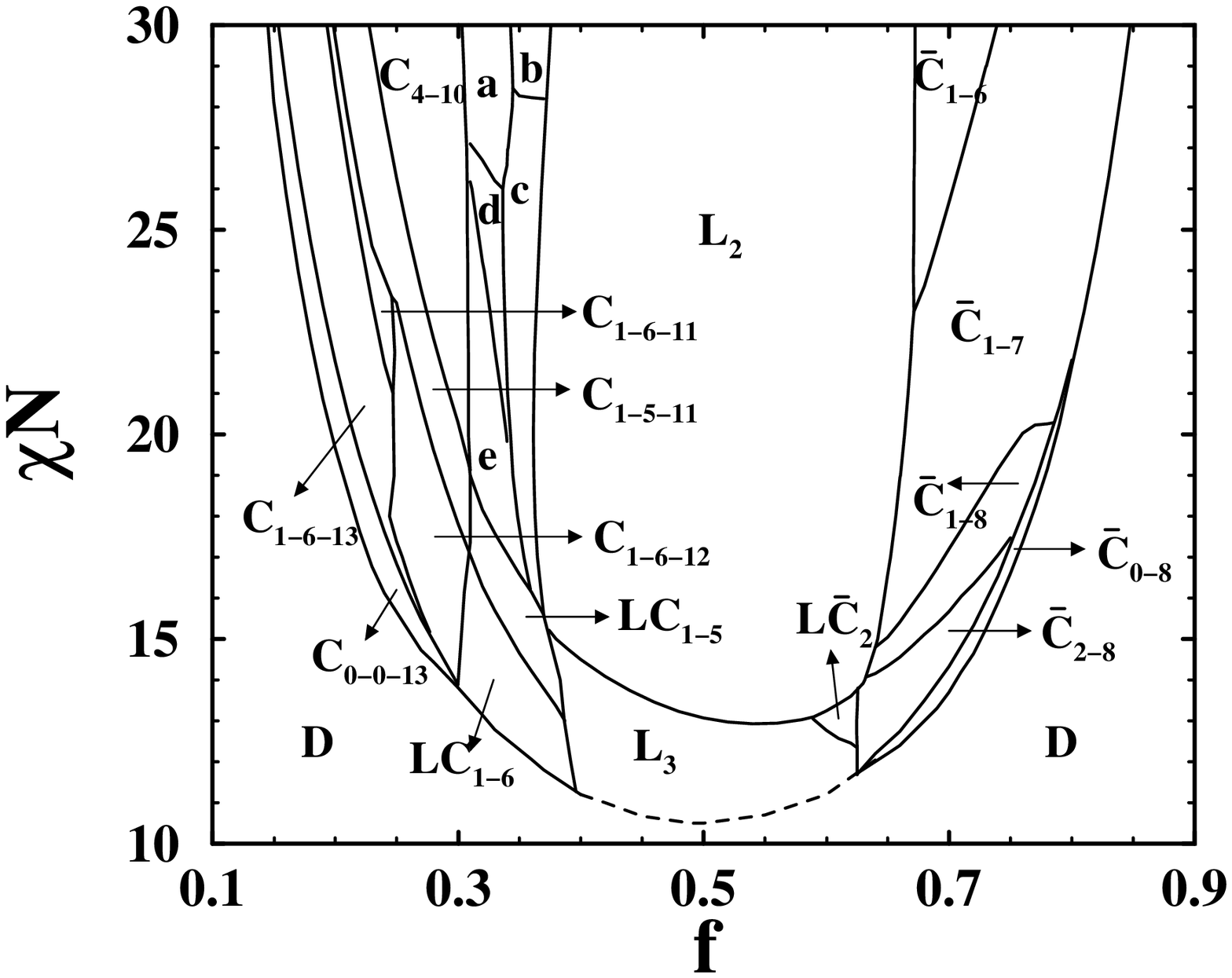, width=0.9\linewidth}
\\ {\bf Figure 2 --- Li}
\end{figure}
\pagebreak
%
%
\begin{figure}[h]
\vspace{3 in}
\epsfig{file=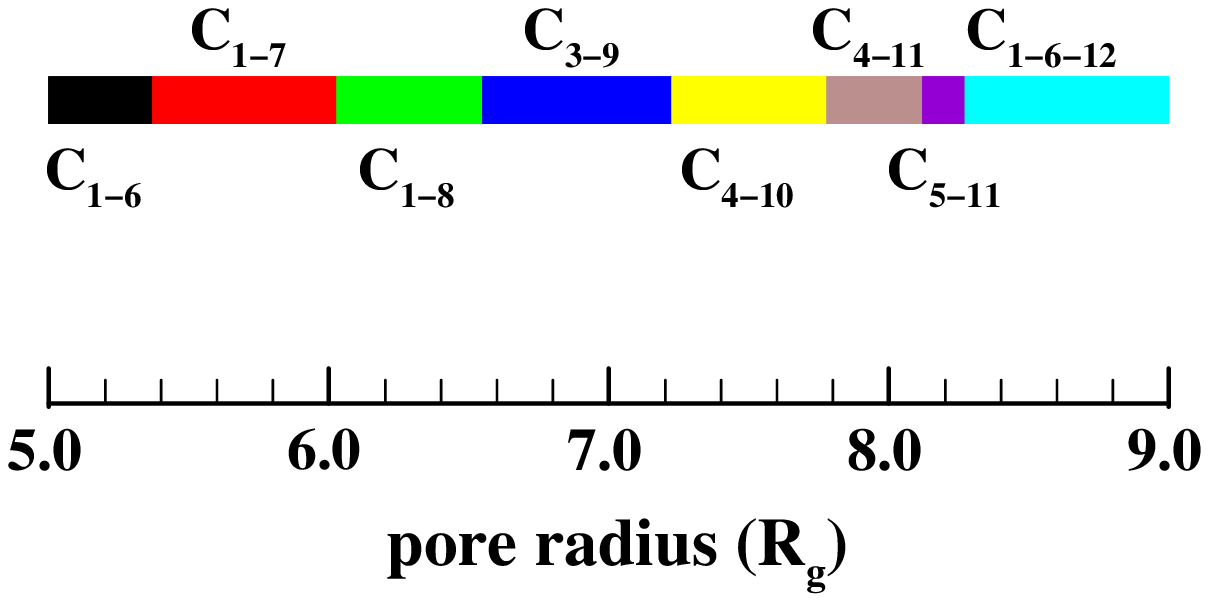, width=0.9\linewidth}
\\ {\bf Figure 3 --- Li}
\end{figure}

\begin{thebibliography}{100}
\bibitem{ACCURACY} While the {\em absolute} free-energies of the structures
are sensitive to the discretization used, 
the {\em differences} between free-energies
are only slightly influenced by discretization. As a consequence, the 
phase boundaries are not significantly changed.
\end{thebibliography}
\end{document}